\documentclass[prb,showpacs,twocolumn]{revtex4}
\usepackage{graphicx}
\usepackage{dcolumn}
\usepackage{bm}
\usepackage{amssymb}
\begin{document}
\def\be{\begin{equation}}
\def\ee{\end{equation}}
\def\ba{\begin{eqnarray}}
\def\ea{\end{eqnarray}}
\newcommand{\s}{\sigma}
\title{Pair distribution function in a two-dimensional
electron gas}

\author{Juana Moreno and D. C. Marinescu}
\affiliation{Department of Physics, Clemson University, Clemson, SC 29634}
\date{\today}
\begin{abstract}

We calculate the pair distribution function, $g(r)$, in a
two-dimensional electron gas and derive a simple analytical
expression for its value at the origin as a function of $r_s$. Our
approach is based on solving the Schr\"{o}dinger equation for the
two-electron wave function in an appropriate effective potential,
leading to results that are in good agreement with Quantum
Monte Carlo data and with the most recent numerical calculations
of $g(0)$. [C. Bulutay and B. Tanatar, Phys. Rev. B {\bf 65},
195116 (2002)] We also show that the spin-up spin-down correlation
function at the origin, $g_{\uparrow \downarrow}(0)$, is mainly
independent of the degree of spin polarization of the electronic
system.

\end{abstract}

\maketitle

\section{\bf Introduction}

There has recently been a growth of interest in studying the pair
distribution function, $g(r)$, in electron gas
models,\cite{davoudi02,capurro02,gori01,Polini01} caused mainly by its
relevance in non-local density functional theories.
\cite{perdew96,chacon88,Gunnar79} The zero inter-electronic
distance value, $g(r=0)$, also appears in the large wavevector and
the high frequency limits of the electronic charge and spin
response functions. \cite{niklasson,zhu84} The importance of
$g(r)$ lies in its connection with the electronic exchange and
correlation of the electron gas model. Moreover, theoretical
calculations of the pair distribution function can be directly
compared with material properties since $g(r)$ is the Fourier
transform of the static structure factor.

The pair-distribution function is  the probability of finding a
pair of electrons at a distance $r$ from each other. Therefore,
the average number of electrons in a spherical shell centered on
an given electron is $n g(r) \Omega_D r^{D-1} dr$, where $\Omega_D
r^{D-1} dr$ is the volume of the D-dimensional shell and $n=N/V$
is the uniform electron density. At large distances, $g(r)$
approaches $1$, whereas near the origin, where the electron charge
is depleted, it is small on account of the Pauli exclusion
principle and the exchange and correlation effects associated with
the Coulomb interaction.

The subject of this paper is an analysis of the pair correlation
function dependence on the inter-electronic distance and electron
density in a two dimensional, interacting, spin polarized
electron system.  Calculations of $g(r)$ in the two-dimensional
paramagnetic electron gas have been reported by Freeman
\cite{Freeman83} and Nagano {\it et~al.}\cite{Nagano84} within the
ladder approximation, by Tanatar and Ceperley\cite{Tan89} using
the diffusion quantum Monte Carlo method (QMC) and, more recently,
by Bulutay and Tanatar\cite{Bulu02} using the hypernetted-chain
approximation (CHNC). Moreover, an analytical expression of $g(0)$
has previously been derived by Polini {\it et~al.}\cite{Polini01}

In order to calculate $g(r)$, we follow the approach developed in
three dimensional systems by Overhauser \cite{Over71} and further
refined in Ref.~\onlinecite{Over95} and \onlinecite{gori01}. This
method is based on the relation between $g(r)$ and the
two-electron scattering problem in an appropriately chosen
effective potential which will be discussed in detail in the
following sections. In addition to obtaining the variation of the
pair distribution function as a function of the coupling-strength,
$r_s$, and of the spin polarization, we also derive an analytic
expression for $g(0)$: 
\be g(0)=\frac{1}{2}\frac{1}{[1+0.6032~ r_s
+0.07263~ r_s^2]^2}. 
\label{eq:corr2nd} 
\ee 
This expression is
found to agree very well with the results of the most recent
numerical calculations.\cite{Tan89,Bulu02} We also compare our
results with the expression derived in Ref.~\onlinecite{Polini01}.

\section{\bf Effective model for the pair distribution function}

A spin-polarized electron gas is characterized, in equilibrium, by
two parameters: the electronic density, or its equivalent $r_s$,
and the polarization, $\zeta = (n_{\uparrow}-
n_{\downarrow})/(n_{\uparrow} + n_{\downarrow})$, where
$n_{\uparrow}$ and $n_{\downarrow}$ are the spin-up and -down
electron densities. For this system, the pair distribution
function is given by \be g(\rho)=\frac{1}{4}[(1+\zeta)^2
g_{\uparrow \uparrow}(\rho)+ 2 (1-\zeta^2)g_{\uparrow
\downarrow}(\rho)+ (1-\zeta)^2 g_{\downarrow \downarrow}(\rho)],
\ee where $g_{\sigma \sigma^{'}}$ are the spin resolved pair
distribution functions. Following Overhauser,\cite{Over95}
$g_{\sigma \sigma^{'}}$  can be related with the two-electron wave
functions as:\cite{gori01} 
\ba 
g_{\uparrow \downarrow}(\rho)=
\frac{1}{2} \langle |\Psi_{\rm singlet}(\rho)|^2
\rangle_{p_{\uparrow \downarrow}(k)} + \frac{1}{2} \langle
|\Psi_{\rm triplet}(\rho)|^2 \rangle_{p_{\uparrow \downarrow}(k)},
\ea \ba g_{\uparrow \uparrow}(\rho)=\langle
|\Psi_{\rm triplet}(\rho)|^2 \rangle_{p_{\uparrow \uparrow}(k)}, \\
g_{\downarrow \downarrow}(\rho)=\langle
|\Psi_{\rm triplet}(\rho)|^2 \rangle_{p_{\downarrow \downarrow}(k)},
\ea
where $\Psi_{\rm singlet}(\rho)$ and $\Psi_{\rm triplet}(\rho)$
are, respectively, the two-electron  wave
function for the singlet and triplet states
and $\langle ...\rangle_{p_{\sigma \sigma^{'}}(k)}$
denotes the average over the probability of finding
two electrons with relative momentum $k$ and spins $\sigma$
and $\sigma^{'}$.\cite{gori01}

The wave function of an electron pair,
${\bf \Psi(\rho)}$, verifies an effective Schr\"odinger equation:
\be
-\frac{\hbar^2}{2 m^*} \Big ( \frac{\partial^2{\bf \Psi} }
{\partial\rho^2}+\frac{1}{\rho} \frac{\partial {\bf \Psi}}
{\partial \rho}+ \frac{1}{\rho^2}\frac{\partial^2 {\bf \Psi}}
{\partial \phi^2} \Big ) + V(\rho)
{\bf \Psi} =E {\bf \Psi},
\ee
where  $V(\rho)$ is the effective
potential,  $m^*=m/2$ is the reduced mass and $E$ is the energy
of the electron pair, which
is approximated by $\hbar^2 k^2 /(2 m^*)$. Since
the solution to this equation can be written as
$\displaystyle {\bf \Psi} = \sum_m \cos{(m \phi)} \Psi_m(\rho)$,
the spin resolved pair distribution functions become
\ba
\label{eq:gupdown}
g_{\uparrow \downarrow}(\rho)= \langle
|\Psi_0(\rho)|^2 \rangle_{p_{\uparrow \downarrow}(k)}
+ 2 \sum_{m=1}^{\infty}\langle
|\Psi_m(\rho)|^2 \rangle_{p_{\uparrow \downarrow}(k)},\\
g_{\uparrow \uparrow}(\rho)=4 \sum_{\rm m \; odd}\langle
|\Psi_m(\rho)|^2 \rangle_{p_{\uparrow \uparrow}(k)}.
\label{eq:gupup}
\ea

Overhauser's method relies on the appropriate selection of an
effective potential capturing the short range correlation effects
of the Coulomb interaction. In three dimensions, Overhauser chose
the electrical potential created by an electron and a neutralizing
sphere of uniform charge with radius $r_s$ surrounding
it.\cite{Over71,Over95} 
The effective potential is expected to
mimic the true one when the relative distance between electrons
verifies $r< r_s$. When $r> r_s$ the potential vanishes and is not
expected to be close to the true potential felt by an electron
moving in a uniform electron gas.
This approach is equivalent to assume   that the
probability of finding three electrons in a sphere of radius $r_s$
is exactly zero.\cite{gori01} Numerical estimates of this
probability for a three-dimensional interacting electron gas\cite{Ziesche00}
have shown that is indeed small
and we expect the same result holds in two dimensions.

Following this procedure, in two dimensions, we might approximate
the screened Coulomb potential by the potential of an electron
surrounded by a circle of radius $r_s$ uniformly filled with
screening charge density $n e =e/(\pi r_s^2)$. For convenience, we
introduce dimensionless variables, $x=\rho /r_s$ and
$\displaystyle V(x)=V(\rho)/(e^2/r_s)$, where $r_s$ is measured in
units of Bohr radius ($a_B=\hbar^2/m e^2$), \ba \label{eq:pot0}
&&V(x)=\frac{1}{x} -\frac{4}{\pi} E(x) ,
\hspace{1in} x \leqslant 1 \\
&&V(x)=\frac{1}{x}
-\frac{ 4}{\pi}x
\Big[
E\Big( \frac{1}{x} \Big) - \Big(1 -\frac{1}{x^2} \Big)
K\Big( \frac{1}{x} \Big) \Big]
, \hspace{0.1in} x \geqslant 1 \nonumber
\ea
where $K(x)$ and $E(x)$ are, respectively,  the complete elliptic
integral of first and second kind.
The screened potential of
a uniformly charged disk of
radius $r_s$ with an electron at its center does not vanish,
but it has an attractive long-range tail,
$\displaystyle V(x \rightarrow \infty)
\rightarrow -\frac{1}{8 x^3}$.
Since we are interested on obtaining an analytical
expression for $g(0)$,
a further simplification of the effective potential
is needed. Since Overhauser's effective potential
is not reliable outside the disk of radius $r_s$,
the most reasonable simplification is to 
make it zero outside this disk.
To avoid a discontinuity in the effective potential
and considering that $V(x)$ is arbitrary  to the
extent that a constant can be added to it, we subtract
from the potential in the region where
$x \leqslant 1$ its value at $ x=1$,
$V_0 =1-(4/\pi)$. Thus,
our effective potential is:
\ba
\label{eq:poteff}
&&V(x)= \frac{1}{x} -\frac{4}{\pi}
E(x) + \frac{4}{\pi} -1,
\hspace{0.1in} x \leqslant 1 \\
&&V(x)= 0
, \hspace{0.1in} x \geqslant 1. \nonumber
\ea

\begin{figure}
\begin{minipage}[t]{\linewidth}
\includegraphics[width=\textwidth]{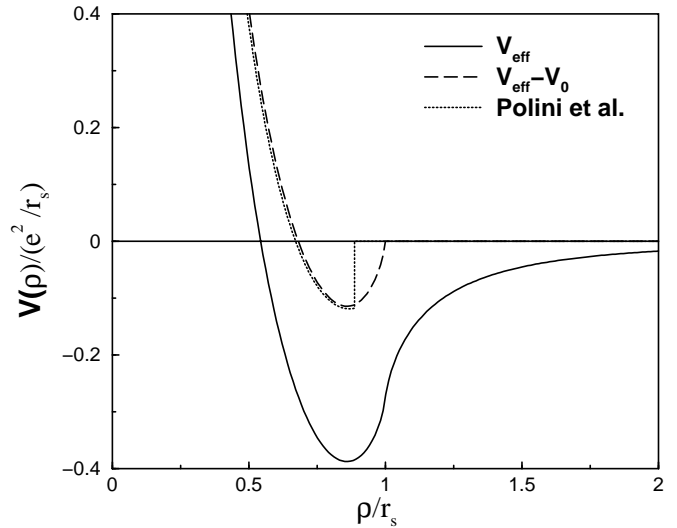}
\end{minipage}
\caption {Normalized effective potential,
$V(\rho)/(e^2/r_s)$, as function of $\rho/r_s$. The potential
from Eq.~(\ref{eq:pot0}) (solid line) and our choice of
effective potential, Eq.~(\ref{eq:poteff}), (dashed line) are displayed.
The effective potential used by Polini {\it et al.}\cite{Polini01}
is also shown (dotted line).}
\label{fig:potential}
\end{figure}

Figure~\ref{fig:potential} displays the initial effective potential
from Eq.~(\ref{eq:pot0}) together with our election of effective
potential, Eq.~(\ref{eq:poteff}), and Polini {\it et al.}
choice \cite{Polini01}, which was based on a previous
variational calculation. \cite{Nagy99}
The main difference between the effective potential
used in Ref.~\onlinecite{Polini01} and ours is that
the former one  has a discontinuity at $\rho=\sqrt{\pi} r_s/2$
while ours is always continuous.

Using Eq.~(\ref{eq:poteff}) for the electronic potential,
the  Schr\"odinger equation becomes:
\ba
\frac{d^2 \Psi_m}{d x^2}+\frac{1}{x}\frac{d \Psi_m}{d x} +
\left( q^2-\frac{m^2}{x^2}\right) \Psi_m(x)=0
\hspace{0.05in},\hspace{0.05in} x \geqslant 1, \nonumber\\
\frac{d^2 \Psi_m}{d x^2}+\frac{1}{x}\frac{d \Psi_m}{d x} +
\left(
q^2-\frac{m^2}{x^2}\right) \Psi_m(x) -\nonumber\\
 -r_s \left( \frac{1}{x} -\frac{4}{\pi}
E(x) + \frac{4}{\pi} -1 \right)  
\Psi_m(x)=0,
\hspace{.1in} \ x \leqslant 1,
\label{eq:Sch}
\ea
where the relative momentum is also renormalized,
$q= k r_s$. The general solution for $x \geqslant 1$ is given by
$\Psi_m(x)= J_m (q x) + B_m(q,r_s) N_m (q x)$, where
$J_m$ is the Bessel function of order $m$ and $N_m$ is the corresponding
Neumann's function.
The coefficient $B_m(q,r_s)$ can be written
as $B_m(q,r_s)=cot(\delta_m(q,r_s))$,
where $\delta_m(q,r_s)$ is the wave function phase shift
due to the presence of the
scattering potential.\cite{phaseshift}

\begin{figure}
\begin{minipage}[t]{\linewidth}
\includegraphics[width=\textwidth]{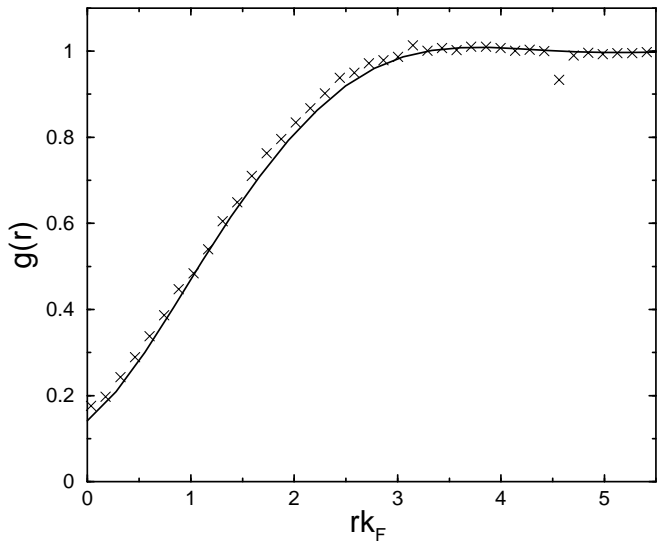}
\end{minipage}
\caption {Pair distribution function of the unpolarized electron
gas at $r_s=1$ as a
function of $r k_F$. Our approximation
(solid line) is compared with the Quantum Monte Carlo data
of Ref.~\onlinecite{Tan89} (crosses).}
\label{fig:gr.rs1}
\end{figure}

\begin{figure}
\begin{minipage}[t]{\linewidth}
\includegraphics[width=\textwidth]{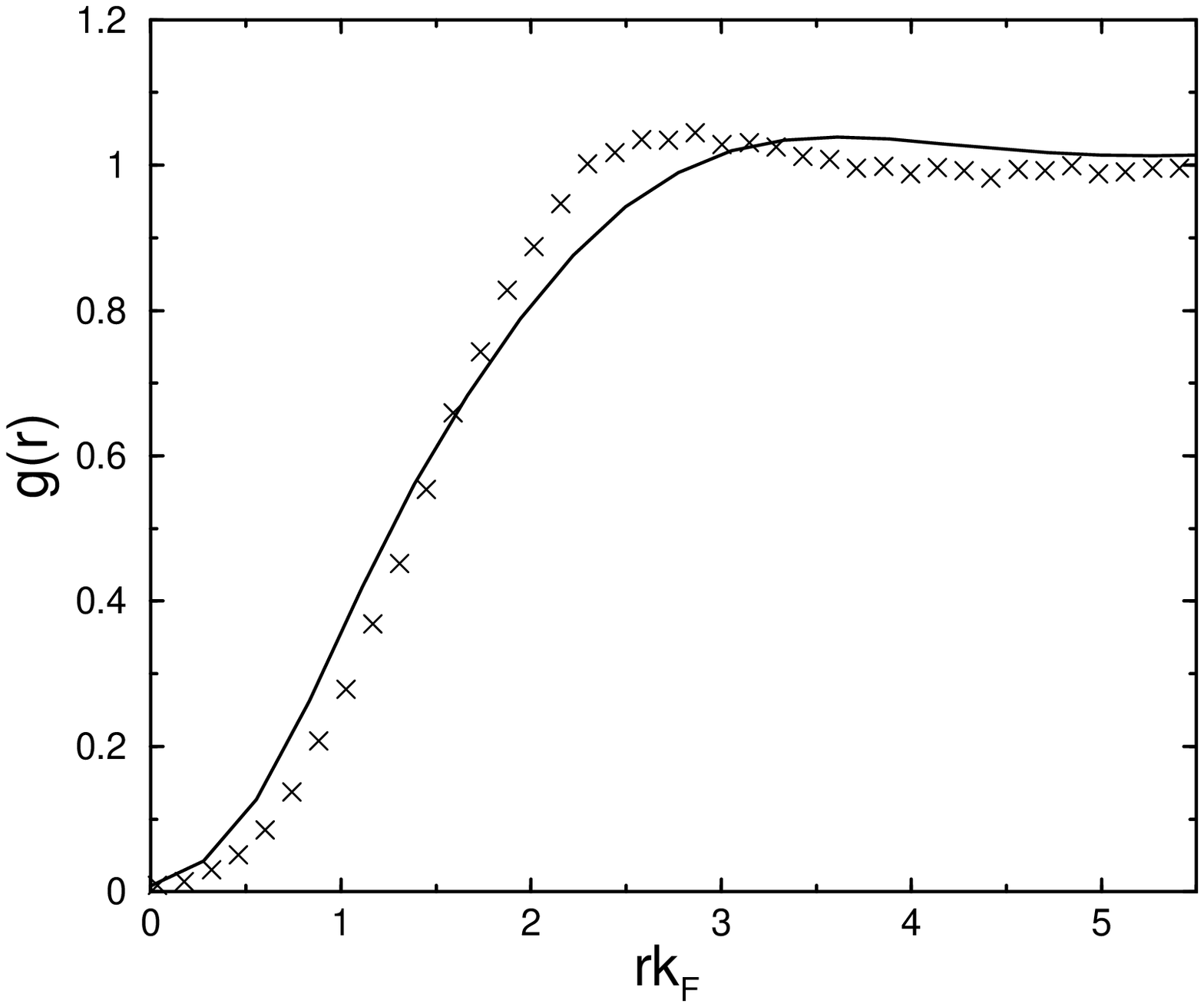}
\end{minipage}
\caption {Pair distribution function of the unpolarized electron
gas at $r_s=5$ as a
function of $r k_F$. Our approximation
(solid line) is compared with the Quantum Monte Carlo data
of Ref.~\onlinecite{Tan89} (crosses).}
\label{fig:gr.rs5}
\end{figure}

To find the solution inside the circle of radius unity
we make a Taylor expansion of the pair wave function:
$\displaystyle \Psi_m(x)=\sum_{n=m}^{\infty} \alpha_{m,n} x^n$.
We arrive to the following recurrent
relation between the  coefficients:
\ba
(n^2 - m^2) \alpha_{m,n} =
r_s \Big\{ \alpha_{m,n-1} + \Big( \frac{4}{\pi}-3 \Big) \alpha_{m,n-2}
\nonumber \\
+ \sum_{r} {\cal A}(r) \alpha_{m,n-2r -2} \Big\}
-q^2 \alpha_{m,n-2},
\label{eq:rec}
\ea
where $\displaystyle {\cal A}(r)=2  \Big[ \frac{(2r-1)!!}{2^r r!} \Big]^2
\frac{1}{2r-1}$.
As a consequence of this recurrent relation,
every $\alpha_{m,n}$ is proportional to $\alpha_{m,m}$ and a function of
$r_s$ and $q$, $\alpha_{m,n}=\alpha_{m,m} F_n(r_s,q)$.

In order to solve Eq.~(\ref{eq:Sch}) we match $\Psi_m(x)$ and its
derivative at $x=1$:
\ba
\alpha_{m,m} G_m(r_s,q)= J_m(q) + B_m(q,r_s) N_m(q) \\
\alpha_{m,m} \widetilde{F}_m(r_s,q)= q  J_m^{'}(q) +
B_m(q,r_s)q  N_m^{'}(q)
\ea
where $\displaystyle G_m(r_s,q)=\sum_{n=m}^{\infty} F_n(r_s,q)$ and
$\displaystyle \widetilde{F}_m(r_s,q)=\sum_{n=m}^{\infty} n F_n(r_s,q)$.
For a given momentum transfer and coupling strength
the parameters $\alpha_{m,m}$ and $\delta_m(q,r_s)$ become
\be
\label{eq:defalpha}
\alpha_{m,m}(q,r_s)=\frac{J_m(q)+ \cot(\delta_m(q,r_s)) N_m(q)}
{G_m(r_s,q)},\hspace{0.05in} {\rm and}
\ee
\be
\cot(\delta_m(q,r_s))=\frac{\widetilde{F}_m(r_s,q) J_m(q)
-G_m(r_s,q)q J^{'}_m(q)}
{G_m(r_s,q)q N^{'}_m(q) - \widetilde{F}_m(r_s,q) N_m(q)}.
\label{eq:defdelta}
\ee

The pair wave functions $\Psi_m(x)$ are computed for any value of
$q$ and $r_s$ using Eqs.~(\ref{eq:defalpha}) and (\ref{eq:defdelta})
and the spin resolved pair
distribution functions are calculated using Eqs.~(\ref{eq:gupdown})
and (\ref{eq:gupup}) and an appropriate choice for the
distribution of the relative momentum of an
electron pair. For simplicity, we use the probability
distribution of a free Fermi gas. For the unpolarized electron
system, the probability of a pair with momentum $q$
is independent of the spin orientation and proportional
to the overlap between
two circles of radius $k_F$ displaced  by $2q$,\cite{Ziesche00}
\be
p_{\sigma \sigma^{'}}(q)=
\frac{16 q}{\pi k_F^2} \Big[ \arccos \Big( \frac{q}{k_F} \Big)
-\Big( \frac{q}{k_F} \Big) \sqrt{1-\Big( \frac{q}{k_F} \Big)^2}
\Big].
\label{eq:unpolprob}
\ee

Figures~\ref{fig:gr.rs1} and \ref{fig:gr.rs5} display our
results for the pair distribution function of the unpolarized system
at $r_s=1$ and $r_s=5$,
respectively. We have used seven ($m_{max}=7$) partial waves and up to
$n=50$ terms for the expansion of $\Psi_m(x)$ in the internal
disk.  Our results at moderate
coupling strengths agree quite well with
the QMC data.\cite{Tan89} However, at larger values of $r_s$
our method is unable to reproduce the strong quantum oscillations
of  the numerical results. 
This discrepancy  is expected and shared by previous calculations
using a self-consistent Hartree scheme.\cite{davoudi02} 
With decreasing dimensionality the role of exchange and correlations 
becomes more important and a screened Coulomb potential
is insufficient to completely capture this physics.
The introduction of self-consistent 
spin-dependent effective potentials have probed able to 
reproduce more closely the
numerical results in this range of densities.\cite{capurro02}

\section{\bf Pair distribution function at the origin}

\begin{figure*}
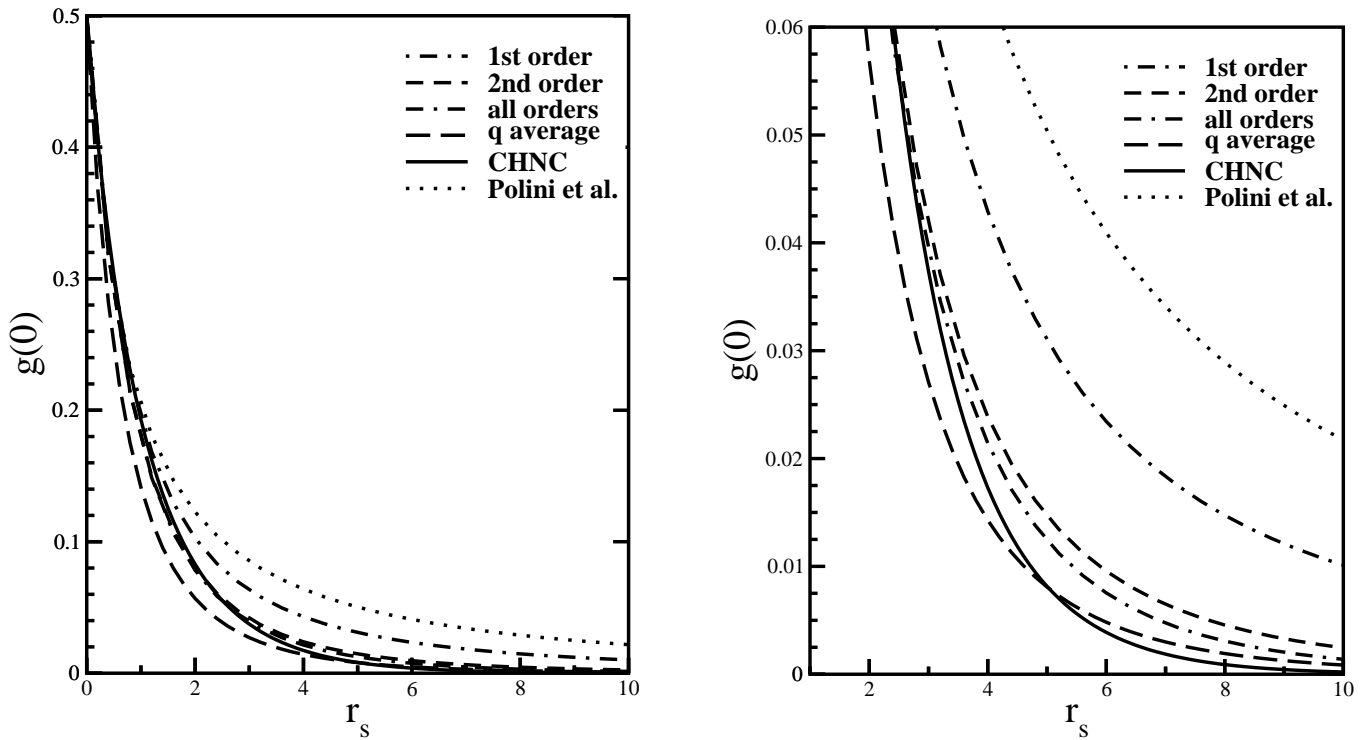

\begin{minipage}[t]{0.47\linewidth}
\includegraphics[width=\textwidth]{corrzero.eps}
\end{minipage}
\hfill
\begin{minipage}[t]{0.47\linewidth}
\includegraphics[width=\textwidth]{smallcorrzero.eps}
\end{minipage}
\caption {Two-particle distribution function at
the origin as a function of $r_s$.
Results of the analytical expansion   to first order given
by Eq.~(\ref{eq:corr1st}) (dot-dashed line),
second order by Eq.~(\ref{eq:corr2nd}) (dashed line),
and infinite order [Eq.~(\ref{eq:correq})] (dot-double-dashed line), and
the average over the distribution of
relative momentum [Eq.~(\ref{eq:qaverage})] (long-dashed line)
are displayed. The numerical results of
Bulutay and Tanatar \cite{Bulu02} (solid line) and
the interpolation results of Polini {\it et al.} \cite{Polini01}
(dotted line) are also displayed. The right panel is a blow-up of the
large $r_s$ region.}
\label{fig:corr}
\end{figure*}

At zero distance,
$g_{\uparrow \uparrow}(\rho=0)$ vanishes  on account of the
Pauli exclusion principle, while $g_{\uparrow \downarrow}(\rho=0)$ is
determined by the $m=0$ component of the two-body wave function,
\be
g_{\uparrow \downarrow}(\rho=0)= \langle
|\Psi_0(\rho=0)|^2 \rangle_{p_{\uparrow \downarrow}(k)}=
\langle |\alpha_{0,0}|^2 \rangle_{p_{\uparrow \downarrow}(k)}.
\ee

Since the distribution of the relative momentum  of an
electron pair is a smooth function, a good estimate of
$\langle |\alpha_{0,0}|^2 \rangle_{p_{\uparrow \downarrow}(k)}$
is obtained by making an expansion around the momentum
where the distribution reach its maximum as:
\be
\alpha_{0,0}\sim \frac{1}{G(r_s)},
\ee
where the momentum dependence of $G$ have been dropped.
Using the recurrent relation~(\ref{eq:rec}), we obtain
a series expansion of $G(r_s)$:
\ba
&&G(r_s)= 1 + r_s \Big\{a_1 +
\sum_{m=1}^{\infty} \frac{{\cal A} (m)}{(2m+2)^2}\Big\}+ \nonumber \\
&&+ r^2_s \Big\{a_2 +\sum_{m=1}^{\infty} {\cal A} (m) \Big[
\frac{1}{(2m+3)^2}\left( 1 +\frac{1}{(2m+2)^2}\right)+\nonumber \\
&&+\left(\frac{1}{\pi}-\frac{3}{4}\right)\frac{1}{(2m+4)^2}
\left(1 + \frac{4}{(2m+2)^2}\right) + \nonumber \\
&&+\sum_{n=1}^{\infty} {\cal A}(n)\frac{1}
{(2n+2)^2 (2n+2m +4)^2} \Big] \Big\}
 +O(r_s^3) \nonumber \\
&&\sim  (1 + 0.6032~ r_s +0.07263~ r_s^2)
\ea
where $\displaystyle a_1=\frac{1}{4} +\frac{1}{\pi}$
and $\displaystyle a_2=\frac{5}{9 \pi} -\frac{1}{6} +
\frac{1}{4} \left(\frac{1}{\pi} -\frac{3}{4}\right)^2$.

We can obtain $g(0)$ at any order in the expansion on
the parameter $r_s$ since that
$\displaystyle g(\rho=0)=\frac{1}{2}g_{\uparrow \downarrow}(0)=
\frac{1}{2} \frac{1}{G(r_s)^2}$.
To first order in the expansion of $G(r_s)$
the pair distribution is:
\be
g(\rho=0)=\frac{1}{2}\frac{1}{[1+0.6032~ r_s]^2}.
\label{eq:corr1st}
\ee
To  second order we recover Eq.~(\ref{eq:corr2nd}).
Our approximation procedure also allows us to
sum all the orders as:
\be
g(\rho=0)=\frac{1}{2} \left( \sum_{n=0}^{\infty}
\frac{\alpha_{0,n}}{\alpha_{0,0}}\right)^{-2}.
\label{eq:correq}
\ee

Finally, we also calculate $g(0)$ performing the average over
the distribution of relative momenta:
\be
g(0)=\frac{1}{2} g_{\uparrow \downarrow}(0)=
\frac{1}{2}
\int_0^{k_F} p(q) |\alpha_{0,0}(q,r_s)|^2 dq
\label{eq:qaverage}
\ee
where $\alpha_{0,0}(q,r_s)$ is given by Eq.~(\ref{eq:defalpha})
and (\ref{eq:defdelta}) and $p(q)$ by Eq.~(\ref{eq:unpolprob}).

Fig.~\ref{fig:corr} displays the pair
distribution function of a two-dimensional unpolarized electron gas
at the origin as a function of the coupling-strength $r_s$.
Results of the first order in the analytic expansion on
$r_s$, Eq.~(\ref{eq:corr1st}), the second order,
Eq.~(\ref{eq:corr2nd}), and the infinite order solution,
Eq.~(\ref{eq:correq}), together with the 
momentum average results, Eq.~(\ref{eq:qaverage}), are displayed.
In addition, the results of the
numerical calculation of Bulutay and Tanatar \cite{Bulu02} and the
recent estimate by Polini {\it et al.}\cite{Polini01} are also included
for comparison. Several conclusions can be gathered.
By adding additional terms in the analytical
expansion on $r_s$ we are able to closely approach
the numerical results\cite{Bulu02} in the low-density regime. Note that
Eq.~(\ref{eq:corr2nd}) is already a
reliable analytical expression for $g(0)$.
Fig.~\ref{fig:corr} also shows
that the results of the momentum average approach,
Eq.~(\ref{eq:qaverage}), are slightly  below  Bulutay and Tanatar results
for small $r_s$, but become even closer to the numerical curve
in the low-density regime.
In this regime, the analytical expression obtained
on Ref.~\onlinecite{Polini01} displays
much larger values of  $g(0)$ than
the available numerical data.\cite{Tan89,Bulu02}

The contact value of the
pair distribution function  changes when the electron
gas is polarized. The spin polarization
directly appears on the expression for $g(0)$,
$\displaystyle g(0)=\frac{1-\zeta^2}{2} g_{\uparrow \downarrow}(0)$.
In addition, the polarization
modifies the distribution of momenta of the electron pair and,
as consequence, the value of $g_{\uparrow \downarrow}(0)$.
We calculate the spin resolved pair distribution function as:
\be
g_{\uparrow \downarrow}(\zeta,\rho=0)=
\int_0^{k_+} p_{\zeta}(q) |\alpha_{0,0}(q,r_s)|^2 dq.
\ee
where $p_{\zeta}(q)$ is 
the distribution of relative momentum in the polarized electron gas:
\ba
&&p_{\zeta}(q)=\frac{8q}{\max(k^2_{F \uparrow},k^2_{F \downarrow})} \quad
{\rm (for} \; 0 \leqslant q \leqslant k_-) \nonumber \\
&&p_{\zeta}(q)=\frac{8q}{\pi k^2_{F \uparrow} k^2_{F \downarrow}}
\Big[ k^2_{F \uparrow}\Big(\arccos(x)-x \sqrt{1-x^2}\Big)\nonumber\\
&&+ k^2_{F \downarrow}\Big(\arccos(y)-y \sqrt{1-y^2}\Big)\Big] \nonumber\\
&& \hspace{1in}{\rm (for}\; k_- \leqslant  q \leqslant k_+)
\label{eq:polprob}
\ea
where
$\displaystyle x=\Big( q +
\frac{k^2_{F \uparrow}-k^2_{F \downarrow}} {4 q}\Big)/k_{F \uparrow}$,
$\displaystyle y =\Big( q -
\frac{k^2_{F \uparrow}-k^2_{F \downarrow}} {4 q}\Big)/k_{F \downarrow}$
and $k_-=|k_{F \uparrow}-k_{F \downarrow}|/2$,
$k_+=(k_{F \uparrow}+k_{F \downarrow})/2$. The Fermi momentum for
the spin up (down) population is related with the polarization and
the Fermi momentum of the unpolarized gas ($k_F$) by
$k_{F \uparrow}=k_F \sqrt{1+\zeta}$
and $k_{F \downarrow}=k_F \sqrt{1-\zeta}$.

Our results show that $g_{\uparrow \downarrow}(\zeta,\rho=0)$  is
largely unaffected by the degree of spin
polarization. The difference between
$g_{\uparrow \downarrow}(\zeta=1,0)$ and its unpolarized counterpart
$g_{\uparrow \downarrow}(\zeta=0,0)$ is, at most,
a few percents for any given  value of $r_s$.
The absence of a significant dependence with the spin
polarization  was also found in
previous calculations.\cite{Polini01}
Moreover,
given that the momentum dependence of our results is rather weak,
we do not expect important changes if the free Fermi
momentum distributions, Eq.~(\ref{eq:unpolprob}) and (\ref{eq:polprob}),
are replaced by the interacting ones.

\section{Conclusions}

We have calculated the pair distribution function in a two
dimensional interacting electron gas, following an approach
originally developed in three dimensions by
Overhauser.\cite{Over95} Within this framework, the short range
correlations of the Coulomb interaction are replaced by an
effective potential, and the calculation of $g(\rho)$  is reduced
to solving the corresponding two-electron scattering problem and
averaging over the probability distribution of the momentum of the
electron pair. Our results for $g(r)$ at moderate coupling
strengths agree well with the numerical data.\cite{Tan89} At
larger values of $r_s$, however, this approximation is unable to
reproduce the strong quantum oscillations of the numerical
results.

The analytic expression for $g(0)$ as a function of $r_s$,
Eq.~(\ref{eq:corr2nd}), derived in this context compares very
favorably with the complete solution of the effective potential,
Eq.~(\ref{eq:qaverage}), and with recent numerical
calculations.\cite{Tan89,Bulu02} We believe that the discrepancy
between the present results and the analytical expression obtained
in Ref.~\onlinecite{Polini01} is essentially due to the different
choice of effective potential (see Fig.~\ref{fig:potential}).
Besides, while we have used the same approach for all values of
the electronic density and polarization, Polini {\it et al.} use
an interpolating scheme between the results of a perturbative
expansion at high-density and the Overhauser's treatment of
scattering processes in the low-density limit.

We have also studied the dependence of $g(0)$
with the spin polarization of the electron gas.
We have found that the spin-up spin-down correlation function,
$g_{\uparrow \downarrow}(0)$, is basically independent
of the degree of polarization. Therefore,
the polarization modifies $g(0)$ only  through
its dependence on the density,
$\displaystyle
g(0)=\frac{2 n_{\uparrow}n_{\downarrow}}{n^2}g_{\uparrow \downarrow}(0)$.

Within this approach, further study of
how the  choice of  the effective potential
modifies the pair distribution function can provide valuable
insight into the short range electronic
correlations in real materials.

\newpage

 {\bf Acknowledgments}

We are grateful to Dr. Bulutay and Dr. Tanatar for providing
us with the results of their numerical calculation.
We acknowledge the financial support provided by the
Department of Energy, grant no. DE-FG02-01ER45897.


\begin{thebibliography}{10}
\bibitem{phaseshift} At large distances the two-electron wave
function can be written as:
$\Psi_m(x) \rightarrow \sqrt{2/\pi qx}\; [
\cos(q x -m \pi/2 -\pi/4) + cot(\delta_m(q,r_s)) \sin(qx-m \pi/2 -\pi/4)]=
[1/\sin{\delta_m(q,r_s)}]\sqrt{2/\pi qx} \;
\sin[q x -m \pi/2 -\pi/4+ \delta_m(q,r_s)]$.
\end{thebibliography}
\end{document}